\documentclass[conference]{IEEEtran}
\IEEEoverridecommandlockouts
\usepackage{cite}

\usepackage{subcaption}
\usepackage{amsmath,amssymb,amsfonts}
\usepackage{algorithmic}
\usepackage{graphicx}
\usepackage{subcaption}
\usepackage{multirow}   
\usepackage{textcomp}
\usepackage{xcolor}
\definecolor{color1}{RGB}{245, 245, 245
}
\definecolor{color2}{RGB}{ 	223, 223, 223}
\usepackage{graphicx} 
\usepackage{url}
\usepackage{listings}

\usepackage[frozencache,cachedir=.]{minted}
\usepackage{easyReview}

\usepackage{booktabs}
\usepackage{longtable}

\def\BibTeX{{\rm B\kern-.05em{\sc i\kern-.025em b}\kern-.08em
    T\kern-.1667em\lower.7ex\hbox{E}\kern-.125emX}}
\begin{document}

\title{Who Introduces and Who Fixes? Analyzing Code Quality in Collaborative Student's Projects }


\author{\IEEEauthorblockN{1\textsuperscript{st} Rafael Corsi Ferrão}
\IEEEauthorblockA{\textit{Insper} \\
São Paulo, Brazil \\
{rafael.corsi@insper.edu.br}}
\and
\IEEEauthorblockN{2\textsuperscript{nd} Igor Dos Santos Montagner}
\IEEEauthorblockA{\textit{Insper} \\
São Paulo, Brazil \\
{igorsm1@insper.edu.br} 
}
\and
\IEEEauthorblockN{3\textsuperscript{nd} Roldofo Azevedo}
\IEEEauthorblockA{\textit{Unicamp} \\
Campinas, Brazil \\
{rodolfo@ic.unicamp.br}}
}

\maketitle

\begin{abstract}

In this full research paper, we investigate code quality education by analyzing how errors are introduced and corrected in group projects. Specifically, we identify which students introduce errors, who corrects them, and when these errors occur throughout the project timeline. The study is set within an embedded systems course, where students are exposed to code quality rules both for the C language and more specific guidelines related to embedded systems. This research explores the dynamics of introducing and correcting code quality errors in group projects, focusing on three key questions:
RQ1: What impact does group formation have on code quality?
RQ2: How students interact in the process of correcting code quality issues?
RQ3: When are code quality issues introduced and fixed during project development?
  
We conducted a qualitative study of data from eight individual lab assessments and two group projects over one course cohorts (N=34) of an undergrad embedded systems course. The course integrates code quality evaluation by providing continuous, automated feedback. After the course ended, we cloned the repositories for all assignments and projects and ran a local analysis tool on every commit, storing the results in a database. We then used git blame to attribute each detected issues to a specific student. When needed, we applied clustering techniques to group students and address the research questions.

Our findings indicate that students who contribute the most code also tend to introduce the highest number of issues. Additionally, issue resolution is frequently postponed until the final stages of project development. We observed no direct correlation between the number of issues in individual lab activities and those introduced in group projects. This may be due to the more structured nature of individual labs, which typically result in fewer issues, compared to the open-ended format of group projects, where issue counts are generally higher. Most issues are resolved by the same student who introduced them; in contrast, cross-student fixes—particularly in shared or boilerplate code tend to take longer. Critical issues are usually addressed promptly, while non-critical ones often persist, suggesting that students prioritize functionality over code quality.




\end{abstract}

\begin{IEEEkeywords}
Embedded Systems; Code Quality; Project-Based Learning 
\end{IEEEkeywords}

{\footnotesize \textsuperscript{}
\thanks{ This work has been submitted to the IEEE for possible publication. Copyright may be transferred without notice, after which this version may no longer be accessible}}

\label{sec:introduction}

\section{Introduction}

Code quality is defined by the attributes that emerge after a program’s initial creation, particularly through the analysis, reflection, and improvement of its static source code characteristics. Among the factors affecting code quality, issues detected by static analyzers are especially relevant in both professional and educational contexts \cite{keuning_systematic_2023}. Static analysis examines a program’s text without executing it, providing programmers with feedback on potential failure points or areas for improvement \cite{louridas_static_2006, gomes_overview_2009}. This type of verification can help reduce a project's technical debt, shorten development time, and minimize production costs \cite{krasnerCostPoorSoftware2020}. Studies \cite{cai_software_2023, kruchten_technical_2012} suggest that education can enhance software development, as programmers often lack adequate techniques, and insufficient technical knowledge directly contributes to technical debt.

Code quality can be introduced at various stages in a student’s education \cite{keuning_systematic_2023}, and educators and researchers use multiple approaches to define and evaluate what constitutes high-quality code (e.g., code smells, variable naming, formatting, static analysis). However, existing research predominantly focuses on individual activities such as code correction and refactoring, leaving the collaborative aspects of group work relatively underexplored.

Group projects tend to be more complex than individual tasks because each team member contributes to the final deliverable, resulting in a higher overall expected workload. Beyond the inherent challenges of collaboration, students in software assessments must manage not only their own code but also code written by their peers, reflecting real-world software development scenarios \cite{berrezueta-guzman_assessing_2025}. In these environments, the need for high-quality code becomes paramount: poor-quality code can directly undermine group performance and impede overall project progress.

Our study examines the dynamics of code quality in group projects within an embedded systems course. Because embedded systems often operate in critical environments, adhering to high-quality coding practices is vital for ensuring correctness and reliability. Despite this recognized importance, research on teaching code quality in embedded systems remains sparse \cite{grimheden_what_2005}, even though guidelines such as MISRA-C are recommended for classroom integration \cite{pasricha_embedded_2022a}.

Most research on code quality focuses on individual work. To the best of our knowledge, no study has specifically examined code quality within the context of student group projects, nor explored the internal team dynamics that arise when students must deliver both a fully functional project and meet additional quality requirements.

To address this gap, we defined the following research questions:

\begin{itemize}
    \item RQ1: What impact does group formation have on code quality?
    \item RQ2: How do students interact in the process of correcting code quality issues? 
    \item RQ3: When are code quality issues introduced and fixed during project development?
\end{itemize}

\section{Related Work}

Research in real projects indicates that early detection of code quality issues minimizes their impact on project development \cite{damm_faultsslipthrough_2006, baca_evaluating_2008, shull_what_2002}. Ayewah et al. \cite{ayewah_google_2010} review Google’s FindBugs project, where engineers analyze errors using an open‐source tool \cite{ayewah_using_2008} capable of detecting Java code defects; in their study, over 77\% of reviews identified genuine issues and produced corrective recommendations.
Kim et al. \cite{kim_prioritizing_2007} evaluate the impact of code quality issues detected by a tool in two open‐source projects, classifying each error’s duration throughout project evolution and its effect on the final product, and errors that can impact code execution are the one correct fasters, when compared with bad practice and style. In a related study from the same author revealing that only a minority of high-priority alerts (between 3\% and 12\%) are intentionally corrected by developers, while 90\% of the remaining errors are removed incidentally through other changes.



In a preliminary study, Scott et al. \cite{scott_reliability_2015} analyze how feedback varies with the experience level of student monitors or teaching assistants. Brown et al. \cite{brown_investigating_2014} observe that instructors lack consensus regarding the severity and frequency of programming errors in a Java course, contrasting actual error rates in student code with instructors’ perceptions. Vanderwerf et al. \cite{vanderwerf_teachers_2024} examine how twelve programming instructors interpret variable naming and its relation to code quality, concluding that there is no common standard in classroom approaches and that perceptions of good naming vary. Many educators rely on manual code quality verification, yet this process is labor-intensive \cite{cook_improving_2022}, subject to reviewer bias \cite{jia_teaching_2022}, and prone to human interpretation errors \cite{corsiferrao_embeddedcheck_2024}.

To scale verification and mitigate these issues, automated tools are widely adopted. Educators and researchers use both general-purpose code quality tools and custom-developed solutions \cite{tran_overcoming_2024}. Deruvo et al. \cite{deruvo_understanding_2018} automatically analyzed 19,000 student codes with a proprietary tool that identified 16 semantic errors via Java’s abstract syntax tree (AST), highlighting common student mistakes. More recently, Rechtackova et al. \cite{rechtackova_catalog_2024a} compared various static analyzers as Flake8, PyTA, Edulint, and Clonediger, and their capacities for detecting code quality defects, with Edulint \cite{rechtackova_giraffereversed_2024} being particularly notable for providing targeted feedback to novice programmers.

An initial strategy for addressing code quality is to offer performance feedback to students, which can motivate them to produce higher-quality code \cite{gomes_software_2021}. However, incorporating additional error messages may yield unintended side effects. Dornbusch et al. \cite{dornbusch_beginning_2024} investigate the relationship between error messages and students’ sense of belonging, finding that such messages can affect students either positively or negatively, depending on the individual. Although this study does not directly link to code quality, it underscores the importance of the manner in which feedback is delivered and how each person is different affected by it.

Regarding student group work and code quality, Jose et al. \cite{josedefaria_forming_2006} propose a group formation strategy based on students’ programming styles. Their findings indicate that more heterogeneous groups show the greatest overall progress throughout the activities, but also face the greatest initial difficulties.

Code quality directly impacts embedded systems, where high-quality code is essential, since these systems typically perform critical functions and are difficult to update after deployment \cite{kim_practice_2009}. Poor code quality can lead to various types of failures that are especially detrimental in embedded systems due to their integration with the real world and the critical nature of their applications \cite{oliveira_investigation_2013}. Motogna \cite{motogna_empirical_2023} conducted an empirical evaluation and noted that the particularities of this domain require approaches distinct from conventional methods. According to the study, in embedded systems, code quality directly influences characteristics such as correctness, security \cite{papadopoulos_interrelations_2018, feitosa_investigating_2015, bianco_tool_2024} and energy efficiency \cite{vetro_deﬁnition_2013}, without necessarily affecting traditional software engineering metrics like maintainability and reusability. 

\section{Context}

This study was carried out at [anonymized], a small private university located in South America, in a embedded systems course offered in the fifth semester (out of ten) in a sequential curriculum of a computer engineering course. The course spans 17 weeks and includes two in-person sessions per week, each lasting 2 hours. The course aims to teach students how to program microcontrollers in the C languange and use their peripherals to solve real-world problems. Students are required to design and develop prototypes that involve software and hardware.

The embedded systems course is structured around two student-centered activities: labs and projects. Labs are activities where students are introduced to new concepts by solving concise, well-defined problems. These problems cover various topics such as reading documentation, working with interrupts, using internal and external microcontroller peripherals, and working with real-time operating systems. These activities are formative and help learn the fundamentals of embedded systems. There is typically one lab per week, all of which are individual assignments, totaling eight labs throughout the course.

Projects are group activities (two students) that are designed to challenge students by encouraging them to apply the concepts they have learned in labs to solve more complex and comprehensive problems. These projects require the application of concepts acquired from multiple labs to develop a unified solution. Projects are the most demanding aspect of the course: students spend part of their class time working on the project, but most of their efforts take place outside the class, designing, coding, and testing their projects.

\section{Code Quality}

In this study, students are exposed to code quality rules on two levels: those pertaining to the C language and a set of course-specific guidelines that address more general aspects of embedded systems, particularly focusing on bare-metal development. For the C language, we use the \textit{cppcheck} static analyzer and all its rules.

Bare-metal code can be understood as code without any intermediate layer between the application and the microcontroller. Due to this direct interaction between the C code and the hardware, certain code quality issues may arise, we check these issues using a custom tool called \textit{[anonymized]}. 

Below, we detail the violations that students use do in the course, which students should avoid.

\subsection{cppcheck}

This study uses the popular \textit{cppcheck} static analyzer \cite{dabruzzopereira_use_2020} to provide C feedback on students' code. It's capable of detecting common mistakes such as memory leaks, null pointer dereferences, out-of-bounds array accesses, and uninitialized variables. By providing detailed warnings and suggestions for improvements, it helps developers proactively address issues and enhance overall code quality.

\subsection{Head File}

This set of rules checks for best practices in structuring C code within \textit{.c} and \textit{.h} files, inspired by the guidelines of MISRA C. The rules emphasize: "Precautions shall be taken to prevent the contents of a header file from being included more than once" and "Separation of Interface and Implementation." Two specific rules have been established to ensure that every header file includes an include guard and that no code is declared in header files, such as function definitions and variable declarations. These rules are named \textbf{noIncludeGuard} and \textbf{cInHeadFile}.

Common violations include students forgetting to implement the include guard correctly in their header files or declaring variables and defining functions within header files.

\subsection{Slow Interrupt Service Routines}

Interrupt Service Routines are functions that are invoked by hardware and can interrupt any other low-level code. It is well-known that we should spend as little time as possible inside an ISR~\cite{koopman2010better}, thus a common mistake is to write slow ISR functions. 
We identify this category of issues by the code \textbf{slowIRS}).

Examples of operations considered violations of this rule include accessing external peripherals (e.g., writing data to an LCD), polling peripheral devices or using any form of loop, invoking software-based delay functions, performing string formatting, or using standard output functions such as {\em printf}.

\subsection{Compiler Optimizations and Volatile}

Compilers apply different optimizations according to the build type. A common compiler optimization is changing how a variable is accessed in memory. This ranges from caching its results in a register to even completely removing a variable. This specific type of optimization can lead to a program having different behavior in \emph{Debug} and \emph{Release} mode. 
Global variables are specially susceptible to these optimizations. Since they are read and written in different functions, it is harder for the compiler to correctly guess when it should apply these optimizations. The \textit{volatile} keyword instructs the compiler to avoid applying optimizations to a variable. Issues related to the incorrect use of {\textit{volatile}} are identified by the \textbf{notVolatileVarIrs} and \textbf{wrongUseOfVolatile}.

Examples of operations considered violations of this rule include updating global variables within an ISR without declaring them as \textit{volatile}, or misusing the keyword by declaring local variables as \textit{volatile}, such as a local variable within the \textit{main} function.

\subsection{Abuse of Global Variables}

Abusing the use of global variables is associated with code quality issues in embedded systems~\cite{kane_toyota_2010}. After the introduction of global variables in ISR functions, some students starts to abuse of global variables indiscriminately.  We refer to issues in this category as \textbf{wrongUseGlobalVar}.

From a best practices perspective, each function should ideally manage its own data or receive it through well-defined interfaces, such as function parameters or return values. By adopting a modular design, it becomes easier to track dependencies, test individual components, and ensure that changes in one area do not inadvertently affect other parts of the program. 

\subsection{Violations example}

The following C code contains examples of violations of certain embedded rules that students commonly make in the dataset we are analyzing, specifically, on this code snippet we have the fallowing issues:

\begin{itemize}
 \item Line 1: \textbf{notVolatileVarIrs} – A global variable accessed by the ISR must be declared as a global variable. 
 \item Line 2: \textbf{wrongUseGlobalVar} – The variable’s scope can be limited to the main function.
 \item Lines 7–8: \textbf{slowIRS} – Functions that slow down the interrupt, such as software delays or sending data via serial communication.
 \item Line 13: \textbf{wrongUseOfVolatile} - This variable does not need to be volatile. 
\end{itemize}

\vspace{2px}

\begin{minted}
[linenos, style=bw, xleftmargin=8pt, numbersep=6pt,
highlightlines={1,2,7,8,13}, highlightcolor=color2] {C}
int flag = 0;
volatile int cnt = 0;

int GPIO_Handler () {
  flag = 1;
  gpio_put(LED, 1);
  sleep_ms(1);
  printf("Debug gpio irs \n");
}

void main (void) {
  ... // Initialization
  volatile int status;
  while (1) {
    if (flag == 1) {
      sprintf(str, "cnt:%d", cnt++);
      // ...
    }
  }
}
\end{minted}

Below is an example of a code snippet with issues detected by cppcheck in a student's code. In this case, the student is generating a square wave on a pin to produce a frequency that will be played on a buzzer:

\begin{itemize} \item line 2: \textbf{zerodivcond} – Possible division by zero if \textit{freq} is zero, which in this code represents silence. \item line 4: \textbf{uninitvar} – The \textit{i} variable in the \textit{for} loop is not initialized. \end{itemize}

\vspace{2px}

\begin{minted}
[linenos, style=bw, xleftmargin=8pt, numbersep=6pt,
highlightlines={2,4}, highlightcolor=color2] {C}
void tone(int freq, int time){
  int periodo = 1000000 / freq;
  int t = freq * time / 1000;
  for(int i; i < t; i++){
    set_buzzer();  
    delay_us(periodo/2);						 
    clear_buzzer();                              
    delay_us(periodo/2);	
  }
}
\end{minted}

\section{Methodology}

In the Fall semester of 2023 semesters, students were introduced to all embedded code quality rules through detailed handouts that combined theoretical explanations with practical exercises. These activities were progressively integrated with the laboratory sessions, and by the fourth laboratory session, all rules had been covered. The cppcheck issues were covered more general, and the tool explained in a similar handout, due to the high number of checks this tool verify, just a small portion of it was covered, but students were exposed to how it works and the importance of it.

A total of 34 students participated in the study, with 8 laboratories each. For the purpose of this analysis, only groups formed by pairs of students were considered, while groups of three or single students were discarded. Specifically, the analysis included 14 pairs for project 1 and 12 pairs for project 2.

All student submissions were made via Git/GitHub, with code quality tools activated on GitHub Actions. Final submissions were only accepted if no issues of any kind were detected by the tools. Additionally, students could automatically check and analyze their code as many times as they wished until the submission deadline.

A post-course analysis was performed, in this analysis, repositories containing individual lab assignments and student projects were cloned, and both tools were executed locally on every student's commit using the tool version that was used during the course. For each assessment and each commit, we collected the following data: all detected issues, the individual responsible for making the changes, the commit message, commit number and the number of changed lines.

Two types of issue metrics were considered when processing the data:

\begin{itemize}
    \item \textbf{Total issues:} The total number of times a given issue appeared across all commits within an assessment.
    \item \textbf{Occurrence of issues:} Counts each issue type only once, regardless of how many times it appeared throughout the commits.
\end{itemize}

The analysis focused on two types of issues:

\begin{itemize}
    \item \textbf{cppcheck}: Code quality issues in the C language, as identified by the \textit{cppcheck} tool.
    \item \textbf{embedded}: Embedded issue related to bare-metal issues, as identified by the \textit{[anonymized]} tool.
\end{itemize}

For the statistical analysis used in this study, we applied the Mann-Whitney U test. This non-parametric test is appropriate for comparing two independent distributions without assuming normality, as verified by visual inspection of all tests in the result section.

%
%

%

\section{Result}

The analysis of student submissions revealed a considerable number of issues across all commits. In terms of total issues, \textit{cppcheck} issues accounted for 2268 instances (83.0\%), while embedded issues comprised 463 instances (17.0\%). When we analyze the occurrence of issues (only counting the first time it appears), \textit{cppcheck} represented 257 instances, and embedded issues made up 99 instances. The occurrence and total number of \textit{embedded} and \textit{cppcheck} issues by category for both project and lab assessments are detailed in Tab. \ref{tab:merged-errors}.

\begin{table}[h]
\centering
\caption{Number of \textit{embedded} and \textit{cppcheck} issues in both project and labs assesment.}

\footnotesize
\begin{tabular}{p{0.6cm} p{1.7cm} cc  cc}\hline
\multirow{2}{*}{\textbf{}} & \multirow{2}{*}{\textbf{Issue}} & \multicolumn{2}{c}{\textbf{Project} (N=26)} & \multicolumn{2}{c}{\textbf{Labs} (N=220)} \\
               &                & \textbf{Occurrence} & \textbf{Total} & \textbf{Occurrence} & \textbf{Total} \\ \hline
\multirow{6}{*}{\rotatebox{90}{\textit{embedded}}} 
               & slowIRS                 & 41 & 123  & 53 & 84 \\
               & wrongUseOfVolatile        & 8  & 29   & 2  & 6 \\
               & wrongUseGlobalVar         & 16 & 78   & 43 & 97 \\
               & cInHeadFile             & 8  & 42   & 13 & 64 \\
               & noIncludeGuard          & 13 & 85   & 0  & 0 \\
               & notVolatileVarIrs       & 5  & 5    & 4  & 13 \\
               \midrule
\multirow{15}{*}{\rotatebox{90}{\textit{cppcheck}}}  
               & constParameterPointer     & 26  & 672  & 57  & 177 \\
               & unreadVariable            & 108 & 577  & 28  & 117 \\
               & invalidPrintfArgType  & 12  & 322  & 29  & 113 \\
               & uninitvar                 & 14  & 283  & 2   & 2   \\
               & variableScope             & 32  & 138  & 163 & 520 \\
               & constVariablePointer      & 7   & 64   & 11  & 58  \\
               & shadowVariable            & 5   & 40   & 16  & 67  \\
               & zerodivcond               & 10  & 34   & 0   & 0   \\
               & unusedVariable            & 7   & 29   & 27  & 75  \\
               & invalidPrintfArgType & 5   & 21   & 38  & 363 \\
               & missingReturn             & 2   & 14   & 2   & 2   \\
               & redundantInitialization   & 1   & 14   & 0   & 0   \\
               & unusedStructMember        & 6   & 11   & 2   & 2   \\
               & legacyUninitvar           & 2   & 8    & 0   & 0   \\
               & constVariable             & 4   & 6    & 0   & 0   \\ \hline
\end{tabular}
\label{tab:merged-errors}
\end{table}

\section*{RQ1: What impact does group formation have on code quality?}

To answer this question, we conducted a series of analyses to determine whether the composition and dynamics of the group can impact the number of issues introduced in the projects.

\subsection{Presence of a dominant student in the project}



To understand whether the difference in students' contribution influences the introduction of issues in projects, we classified the groups into two categories based on each student's contribution to the lines of code (LOC). We considered the percentage of LOC each student contributed to the final submission and categorized the groups as follows:

\begin{itemize} 
    \item \textbf{Cluster 0:} Represents groups where the contribution in LOC is more balanced, ranging between 50\% and 70\% for the student with the highest contribution. This means that both individuals have a similar contribution to the final submission.
    \item \textbf{Cluster 1:} Represents groups in which the student with the highest contribution accounts for 70\% to 100\% of the LOC, meaning that one individual in the group contributed significantly more to the final submission, we call this a dominant student. 
\end{itemize}

The analysis shows that Cluster 0 has a slightly higher average occurrence of issues count at approximately 9.05 (SD = 8.45) per group compared to Cluster 1's average of 7.4 (SD = 5.50). However, the Mann-Whitney U Test yielded a p-value of approximately 0.974, indicating that the difference is not statistically significant. 


When analyzing each cluster individually, we observe the following: In cluster 0, the student with the greater code contribution in terms of LOC introduced an average of 7.14 issues occurrences  (SD = 6.30), while the student with the lower contribution introduced an average of 1.90 issues occurrences (SD = 3.17). In cluster 1, the differences between the two students are much smaller; in this case, the student with the greater contribution added an average of 3.0 issues occurrences (SD = 2.12), and the one with the lower contribution added an average of 4.4 issues occurrences (SD = 4.40). When comparing the values statistically, we could not obtain a statistically significant difference between the number of issues introduced by the two groups.





\subsection{The Impact of Lab Issues on Project Quality}

\begin{figure}[ht!]
    \centering 
    \subfloat[Issues in individual laboratory activities (LAB) and the number of issues inseted in projects  by the same student (Project) (N=34)]{
        \includegraphics[width=1\linewidth]{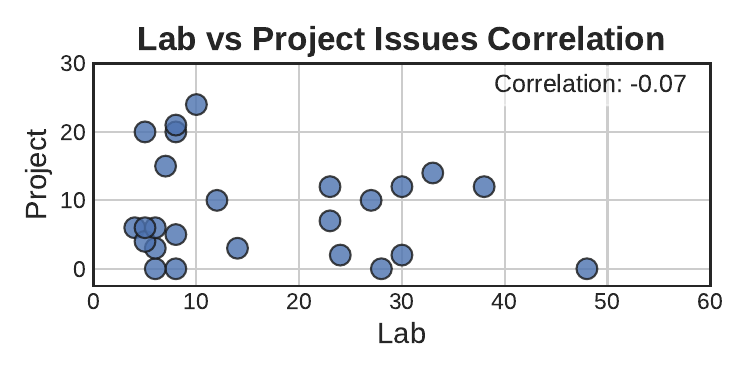}
                \label{fig:lab-errors} 
    }
    \vspace{0.5em} 
    \subfloat[Issues made in the development of each project (N=18)]{
        \includegraphics[width=1\linewidth]{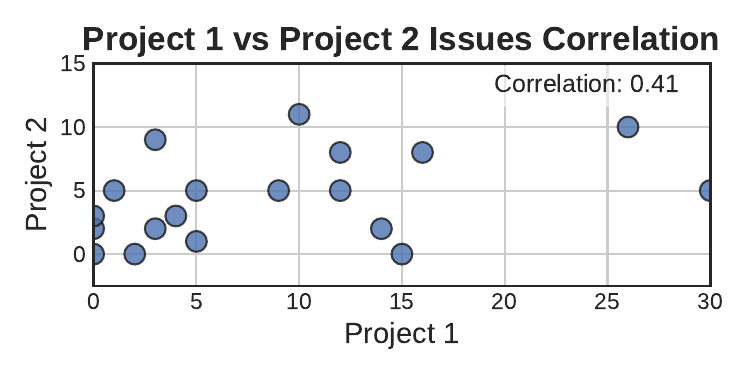}
                \label{fig:project-errors} 
    }
    
    \caption{Analyses of the correction of issue occurrences made by students in the lab, in the project, and in both projects.}
    \label{fig:proj-lab-errors-individual}
\end{figure}

One hypothesis we considered is that the student who makes the most issues in individual laboratory activities is also the one who introduces the most issues in group projects, as they may have unresolved conceptual deficits from the labs. Figure \ref{fig:lab-errors} presents a graph correlating the total number of individual issues made by each student across all laboratory activities with the number of issues they introduced throughout the two projects.

To assess this relationship, we calculated the Pearson correlation coefficient (\textit{r}) between these two variables (individual issues in labs and individual issues in project), obtaining a value of -0.07. This result indicates that there is no correlation between the number of issues made in individual labs and the issues introduced in group projects.

Another idea we can analyze is whether groups formed by students who introduce the most issues in laboratory activities are also the groups with the highest number of issues in projects, different from last analyzes here we use both students' number of issues, not only individual ones. To investigate this, we summed the individual laboratory issues of both students in each group, and we counted the total number of issues introduced by the group in the project and applied the same analysis as before. The resulting Pearson correlation coefficient was r = -0.02, indicating no relation between both variables.

\subsection{Influence of Individual Behavior on Project}

To investigate whether an individual student's behavior influences the number of issues introduced in projects and the violation of code quality rules, we analyzed the number of issues each student introduced in the two different projects (Fig. \ref{fig:project-errors}). We obtained a Pearson correlation of 0.41, indicating a moderate relationship between the issues across projects. This suggests that the behavior of individual students does impact the overall number of issues in the projects.



\subsection{Grades}

The descriptive statistics indicate that the average grade is around 6.48 with issues averaging about 8.76. The scatter plot indicates that most data points are widely spread; for instance, while some of the grade 5 entries are associated with issues as low as 0, others extend up to 37. Although higher grades such as grade 10 appear in cases with fewer issues, the pattern is not consistent across the dataset, suggesting that there is no clear visual separation among the groups. Quantitative analysis indicates that the relationship between grades and the number of issues occurrences is weak. The Pearson correlation coefficient is approximately -0.21, indicating only a slight negative association.

\subsection{Remarks}


We cannot conclude that the presence of a dominant student, who contributes a relatively higher number of lines of code to the project, influences the total number of issues introduced, since our comparison did not reveal any significant difference when compared to groups with a more balanced working dynamic.

Additionally, it is not possible to conclude that individual student performance on lab assessment directly impacts the number of issues, as no correlation was observed between the number of issues a student introduces individually during lab activities and those they introduce in the project. The strongest correlation emerged when comparing the same student across two projects, suggesting a relationship between the student and the number of project issues. This may indicate that external factors, not measured by this study's methodology, have a greater influence on students’ behavior in generating code in a project that is different from labs.


In this study, academic performance alone does not reliably predict the number of issues. We believe that this may be partially explained by the project evaluation format used in the course, as projects are assessed using specification grading. In this model, students can choose from a range of options that may impact their final grade; thus, two groups with an A grade might have delivered projects with different functionalities. Additionally, external parameters, such as the quality of the physical prototype, can affect the final grade, meaning that a higher score does not necessarily correspond to a larger or more comprehensive software project.

\section*{Q2: How do students interact in the process of correcting code quality issues?}

\begin{figure*}
    \centering
    \includegraphics[width=1\linewidth]{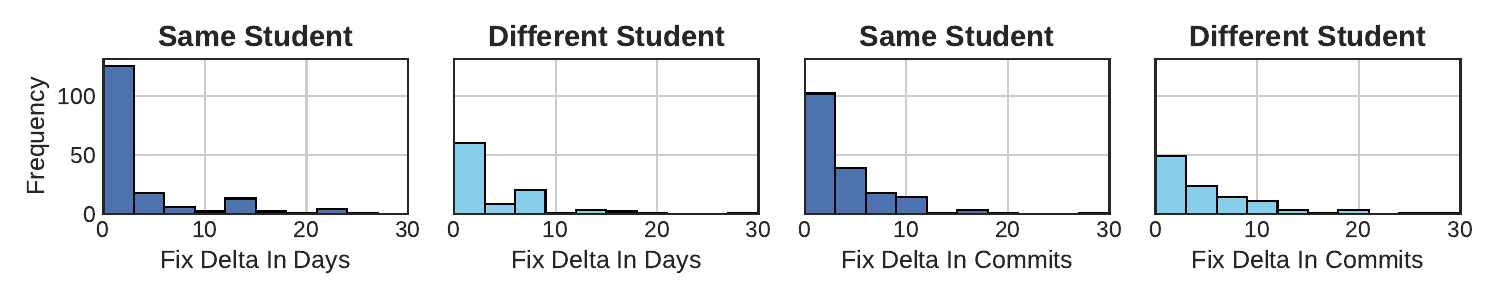}
    \caption{Time taken to resolve a code quality issue by the original student versus a different student, both by commits and by days.}
    \label{fig:fix-histogram}
\end{figure*}

%
%
%

To examine the dynamics of code quality issue correction, we analyzed the commit history to identify which student introduced each issue and who later corrected it, whether it was the same student or his partner. This analysis resulted in {64.37\%} of the issues were fixed by the {same student} who introduced them, while {35.63\%} were corrected by a {different student}. 

Figure~\ref{fig:fix-histogram} shows a histogram of the number of commits and days required to fix a code quality issue. We observed that issues are resolved with fewer commits when fixed by the student who originally introduced them (mean = 3.77, SD = 4.48 commits) compared to when another student corrects them (mean = 6.05, SD = 8.59 commits). This difference is statistically significant (p-value = 0.0057), indicating that the original authors resolved their own issues more promptly. Conducting the same analysis based on the number of days the issue remained active, we obtained an average of 4.60 days (SD = 9.83) when they are fixed by the same student and a higher value of 8.45 days (SD = 14.29) when they are corrected by someone else (p-value = 0.0002).

\subsection{Pre-existing Code Quality Issue}

In the first project, the base code provided to students contained a pre-existing code quality issue (detected by \textit{cppcheck} as type \textbf{uninitvar}). In 12 out of the 14 groups, we were able to trace this issue and identify when it was fixed. In this case, we consider the fix to be associated with the group where the issue was corrected by a different student. 

We observed that these template-originated issues do not follow the same correction pattern as those introduced by the students themselves. For the pre-existing issue, the average number of commits until correction was much higher with a mean of 12 (SD = 8.34) commits, with an extreme case reaching 30 commits only being fixed in the final commit. 

\subsection{Remarks}

Most issues are corrected by the same student who introduced them, which aligns with common group dynamics where the individual responsible for a functionality is expected to refine the code until the issue is resolved, we also need to consider that half of groups were classified in Cluster 0, where a single student domineer the development of the project, inflating this result.

Corrections made by the original author tend to occur more quickly than those addressed by a different student, possibly due to deferring responsibility for tasks outside one's own work or the additional difficulty of modifying someone else's code. This effect is even more pronounced for issues in the provided boilerplate, where corrections take significantly longer, likely because neither student originally developed that section of the code and thus lacks ownership over it.

Similar to findings in the literature analyzing large-scale software projects \cite{kim_prioritizing_2007}, issues with a shorter lifespan are those that are critical and can impede program execution and functionality. For instance, failing to use the keyword \textit{volatile} in a global variable accessed in an interrupt service routine (ISR) can cause the code to malfunction. Conversely, issues that do not affect the program's functionality tend to persist longer within projects. For example, excessive use of global variables may not directly affect program behavior, but impacts code quality parameters.

\section*{RQ3: When are code quality issues introduced and fixed during project development?}

\begin{figure}
    \centering
    \includegraphics[width=1\linewidth]{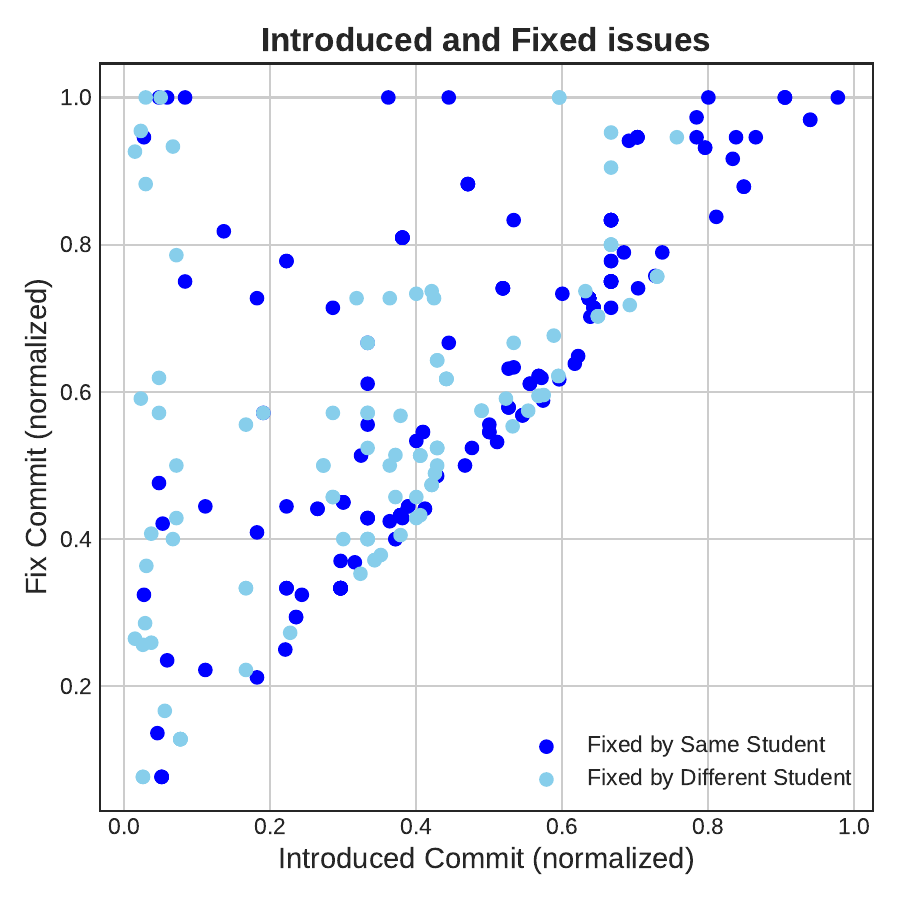}
    \caption{Visualizations showing when code quality issues were introduced and removed by commits and days, comparing fixes by the same student vs. a teammate (N=180).  Normalized commits, where 0 is the first commit and 1 is the last one.}
    \label{fig:intro-removed-commits}
\end{figure}

%
%

Figure~\ref{fig:intro-removed-commits} displays a plot showing when an issue was introduced and when it was removed by number of commits normalized.  On average, issues are introduced later at commit position 0.48 (SD = 0.24) when they are fixed by the same student, compared to 0.35 (SD = 0.22) when a different student fixes them (p-value = 0.0335).  We observed that 14 issues were fixed in the last commit, the majority (70\%) of which were fixed by the same student. 

There is a higher tendency for issues to be introduced by different students in the early stages of the project. However, this trend shifts around the midpoint of development after approximately 0.4 commits on the normalized scale, issues are more likely to be fixed by the same student who introduced them. Notably, the graph shows that by the end of development, all issues are both introduced and fixed by the same student.

When conducting the same analysis based on days, we observed a higher concentration of issue corrections on the last day. In this analysis, we found that 60 issues were corrected on the final day (48 by a single student and 12 by another). Additionally, 14 issues were created on the deadline day itself.

\begin{figure}[ht!]
    \centering
    \includegraphics[width=1\linewidth]{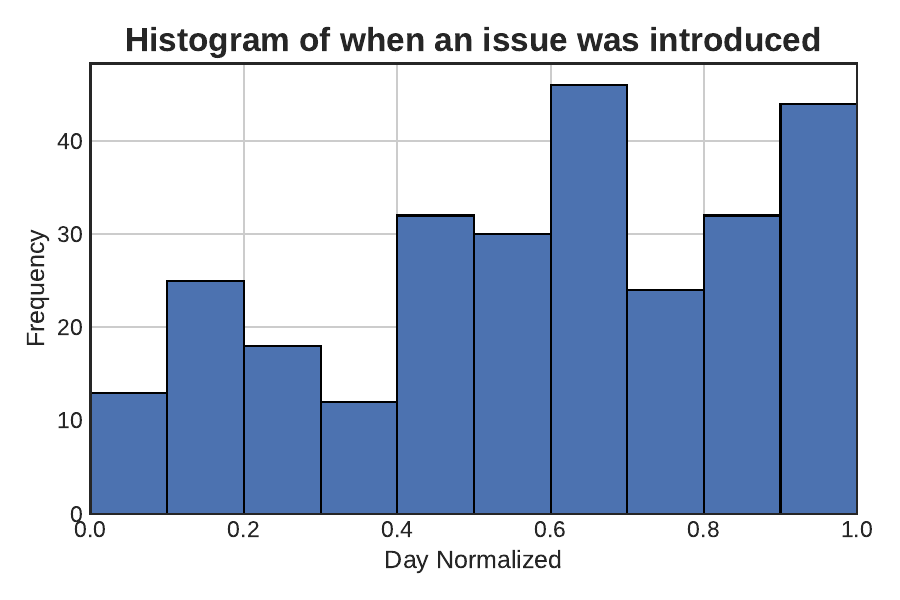}
    \caption{Visualizations showing when an issue is added during code development in days since the beginning of the project (0) and the last day (1).}
    \label{fig:day-histogram}
\end{figure}

Figure \ref{fig:day-histogram} is a histogram depicting when issues were introduced in the project by days nornalized of project development. On average, issues are introduced starting around the middle of the project (mean = 0.59, SD = 0.28). We used the project’s development days as the unit of measurement, with 0 representing the project’s start and 1.0 representing the end. We applied a normalized scale because Project 1 is significantly shorter than Project 2.




\subsection{By Types Of Issues}

\begin{table}[ht]
    \centering
    \caption{Mean and std of number of commits to fix specific types of issues, critical col indicates when an issue is considered critical.}
    \begin{tabular}{llrrrc}
        \textbf{type} & \textbf{Issue} & \textbf{mean} & \textbf{std} & \textbf{count} & \textbf{critical} \\
        \midrule
        embedded & badUseGlobalVar          & 6.00 & 6.20 & 15 &  \\
        cppcheck & constParameterPointer    & 5.00 & 5.66 &  2 &  \\
        embedded & noIncludeGuard           & 4.40 & 3.44 &  5 &  \\   
        cppcheck & zerodivcond              & 1.33 & 0.71 &  9 & YES \\
        cppcheck & syntaxError              & 1.00 & 0.00 &  4 & YES \\
        cppcheck & uninitvar                & 1.00 & 0.00 &  2 & YES \\
        embedded & notVolatileVarIrs        & 1.00 & 0.00 &  5 & YES \\
        cppcheck & constVariablePointer     & 1.00 & 0.00 &  3 &  \\
    \bottomrule
    \label{tab:per:issue}
    \end{tabular}
\end{table}

To examine whether the type of issue \textit{cppcheck} or \textit{embedded} affects how students interact with them, we compared the number of commits made before correction. For \textit{cppcheck} issues, students made an average of 5.09 commits (SD = 7.26) before resolving the issue. In comparison, \textit{embedded} issues required an average of 3.38 commits (SD = 3.61). This difference was not statistically significant (\textit{p} = 0.0810).

Table~\ref{tab:per:issue} details the issues by type and the average time taken to fix each one. We observed that the most time-consuming issue to be corrected by students was an \textit{embedded} issue related to the improper use of a global variable. This issue does not directly affect the dynamic execution of the code, which may explain the delayed correction. On the other hand, the fastest issues to be fixed were mostly critical ones that impact either the dynamic behavior of the program or prevent it from compiling. These include \textbf{zerodivcond}, \textbf{syntaxError}, and \textbf{uninitvar} all detected by \textbf{cppcheck} and related to variables being used before initialization as well as \textbf{notVolatileVarIrs} (\textit{embedded}).


\subsection{Remarks}

The results indicate that students tend to introduce more issues in intermediate commits, particularly from the middle to the end of the project. We observed a peak in issues near the project deadline, suggesting that students become more focused and dedicated as the submission date approaches. Additionally, there is a clear tendency among students to postpone issue correction until the final day. Based on our observations, students prioritize implementing functionality first and only later address code quality issues to ensure their project is accepted, as we do not accept any submissions with issues detected by the tools.


The accidentally introduced issue in the sample code of Project 1 is a critical one, as detected by cppcheck as \textbf{uninitvar} where a variable intended to act as a pointer within an array to transmit a string via serial communication is not initialized to zero, as shown on the snippet below:

\vspace{2px}

\begin{minted}[linenos, style=bw, xleftmargin=8pt, numbersep=6pt,
highlightlines={2}, highlightcolor=color1] {C}
uint32_t usart_puts(uint8_t *pstring) {
  uint32_t i;
  while(*(pstring + i)) {
    ...
  }
}
\end{minted}

Although this issue is significant, the function \textit{usart\_puts} was not used in the code and belonged to the instructor's legacy solution for validation purposes. We believe that its lack of use and the fact that it was not introduced by either student likely contributed to the issue's extended lifespan.

\subsection{Threats to Validity}

A potential threat to validity is selection bias, as this study examined only submissions from a specific embedded systems course at a single university, and all projects were carried out in pairs. We also worked with a limited number of students and groups, and different behaviors may emerge when analyzing a larger sample size or different group lengths.

Another threat is that only issues associated with commits were recorded. This implies that not all issues may have been captured in the analysis, as it depends on students creating a new commit to generate a code snapshot that can be analyzed. There is also uncertainty regarding whether students actually saw or acknowledged the issue messages generated by the tools. The analyses were conducted on GitHub, which requires that students take an active role by submitting their code, waiting for the tools to execute, and, when an issue exists, opening the generated log and reading what the issue was. Therefore, the students need to actively check to see if their code contains any quality errors. We know that they verify their code at least once per submission because they must deliver code that does not violate any quality rules.

Finally, it is unclear whether issues were genuinely corrected by the students or if they simply disappeared due to other modifications in the code. 

\section{Conclusion}

In this study, we found that external factors not directly related to students’ individual performance in the course can influence the number of issues introduced in their projects. Additionally, students tend to view issues as less important, often postponing their correction until close to the deadline. Although individual performance does not directly correlate with the number of issues, comparing the same student across different projects suggests a relationship between their work habits and the frequency of issues.

Most issues are corrected by the same student who introduced them, aligning with common group practices. Corrections by different students take longer, especially when dealing with boilerplate code that lacks clear ownership. Critical issues (e.g., failing to use the \textit{volatile} keyword in an ISR) are fixed more rapidly, whereas non-critical issues (e.g., excessive global variables) often remain, reflecting the importance of functionality over code quality.

Integrating group-based projects that emphasize not only functional outcomes but also code hygiene and adherence to standards offers a significant advantage, as these assignments involve more complex projects where code quality becomes even more critical. Such curricular enhancements could bridge the gap between academic exercises and industry demands, ensuring that students learn to navigate both individual and collective coding responsibilities.

Future work could involve a qualitative analysis of student interactions, examining how they collaborate during project development and how they respond to automated code verification and quality rules. Additionally, this research opens the door to further studies on best practices for managing code quality in group projects. Future studies might explore how different collaborative strategies influence coding standards, the spread of issues, and the overall development process.

\bibliographystyle{IEEEtran}
\bibliography{zotero}

\end{document}